\documentclass{article}
\usepackage{microtype}
\usepackage{graphicx}
\usepackage{booktabs}
\usepackage{multirow}
\usepackage{xspace}
\usepackage{amsmath}
\usepackage{amssymb}
\usepackage{tikz}
\usepackage{hyperref}
\newcommand{\ik}{\textsf{ik}\xspace}
\newcommand{\ikp}{\textsf{ik-P}\xspace}
\newcommand{\Row}{\textsf{Row}\xspace}
\newcommand{\Col}{\textsf{Col}\xspace}
\newcommand{\icl}{Ice~Lake\xspace}
\newcommand{\emr}{Emerald~Rapids\xspace}
\newcommand{\amxp}{\textsf{AMX-prefill}\xspace}
\usepackage[accepted]{mlsys2025}
\newcommand{\arxivauthorblock}{%
\mlsyssetsymbol{equal}{*}
\begin{mlsysauthorlist}
\mlsysauthor{Runguo Li}{equal,uiuc}
\mlsysauthor{Dezhen Lu}{equal,tue}
\end{mlsysauthorlist}
\mlsysaffiliation{uiuc}{University of Illinois Urbana-Champaign, Urbana, IL, USA}
\mlsysaffiliation{tue}{University of T\"{u}bingen, T\"{u}bingen, Germany}
\mlsyscorrespondingauthor{Runguo Li}{runguol2@illinois.edu}
\mlsyscorrespondingauthor{Dezhen Lu}{dezhen.lu@student.uni-tuebingen.de}
}
\newcommand{\arxivpdfauthor}{Runguo Li, Dezhen Lu}   %
\newcommand{\arxivequal}{\mlsysEqualContribution}   %
\newcommand{\arxivnotice}{Preprint.}                 %

\makeatletter\renewcommand{\mlsys@appearing}{\arxivnotice}\makeatother
\mlsystitlerunning{Weight Placement Is Not Enough: Synchronization in Multi-Node CPU Decoding}
\begin{document}
\twocolumn[
\mlsystitle{Weight Placement Is Not Enough:\\Synchronization in Multi-Node CPU Decoding of Ternary LLMs}
\arxivauthorblock
\mlsyskeywords{LLM inference, CPU, NUMA, ternary models, synchronization}
\vskip 0.3in
\begin{abstract}
Ternary LLMs read only a few gigabytes per decoded token, yet on dual-socket servers with four NUMA nodes the fastest ternary runtime we found uses less than half of the memory bandwidth.
Placing each weight on the node of the thread that reads it is necessary but not enough: in a decode engine with node-local weights that reuses ik\_llama.cpp's kernels, threads on two generations of Xeon servers still spend 20--48\,\% of each token waiting at synchronization points.
Per-thread barrier traces show that most of the wait is for the slowest threads of each node; the barriers' own latency (microseconds each, but up to 12\,\% of the token for the smallest model) and serial per-node work are smaller parts.
On identical kernels, a Megatron-style column split beats a row-owned split without cross-node reductions in every configuration we measured, and isolating the number of barriers shows that most of its gain comes from its threads waiting less for each other and for their node leaders, and only a minority from having fewer barriers.
Where decoding is DRAM-bound, the resulting engine is 1.5--2.1$\times$ faster than ik\_llama.cpp with all cores and 1.5--1.9$\times$ faster than ik\_llama.cpp at its fastest thread count (3.2$\times$ and 2.8$\times$ for a small model that runs from the L3 cache), at the same accuracy and with less energy per token.
\end{abstract}

]
\hypersetup{pdftitle={Weight Placement Is Not Enough: Synchronization in Multi-Node CPU Decoding of Ternary LLMs}, pdfauthor={\arxivpdfauthor}, pdfsubject={}, pdfkeywords={LLM inference, CPU, NUMA, ternary models, synchronization}}
\printAffiliationsAndNotice{\arxivequal}

\section{Introduction}\label{sec:intro}

Ternary language models restrict their weights to $\{-1, 0, +1\}$ and store them in about two bits each~\cite{ma2024era}.
An eight-billion-parameter model then reads about 2\,GB per generated token, which brings single-stream decoding within reach of CPU servers, whose memory bandwidth, spread over several sockets and sub-NUMA nodes, adds up to hundreds of GB/s.
On one NUMA node, decoding is close to bandwidth-bound: each token reads all weights once, and the best runtimes use 72--78\,\% of the node's bandwidth.

\begin{figure*}[t]
\centering
\includegraphics[width=\textwidth]{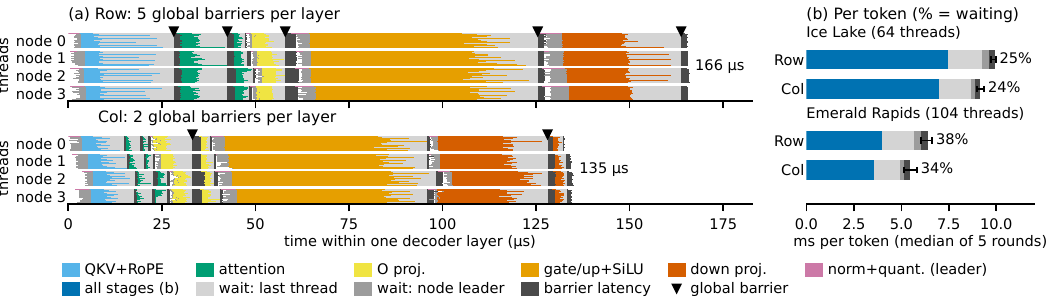}
\vspace{-0.25in}
\caption{What the threads do once every weight is read from its own node. (a) One of 32 decoder layers of the median-duration token of a traced run (Emerald Rapids, Llama3-8B IQ2\_BN, 104 threads on 4 nodes): each row is a thread, colored while computing, grey while waiting for the last thread (of its node, or of all nodes at a global barrier, $\blacktriangledown$), for its node leader, or for the barrier itself. (b) Per-token time split the same way: medians of 5 traced rounds (whiskers: min--max), with the share spent waiting.}
\label{fig:teaser}
\end{figure*}

Using the whole machine is harder.
Across the four NUMA nodes of a dual-socket Ice Lake server, the fastest ternary runtime we found, ik\_llama.cpp (\ik), reads DRAM at only 48\,\% of the machine's local read bandwidth (Llama3-8B); on an Emerald Rapids server it becomes 3--12\,\% \emph{slower} when given 104 instead of 64 cores.
The natural diagnosis is weight placement, and placing every weight row on the node of the thread that reads it is indeed necessary; recent systems such as ArcLight~\cite{xu2026arclight} build on node-local tensor parallelism.

This paper asks what limits decoding \emph{after} weight placement is solved.
We fix placement inside \ik and then replace its execution model by an engine that keeps \ik's kernels and weights but synchronizes only a few times per layer, on two generations of Xeon servers, with three ternary models and the same models in three non-ternary formats.
Time moves from reading weights to waiting at synchronization points (Fig.~\ref{fig:teaser}).
This shift has been observed for weights that fit in GB-scale caches~\cite{zhang2026cacheresident} and for fast GPU memory~\cite{chen2026membound}; we show that it already limits DRAM-bound decoding on standard multi-node servers, decompose the waiting, and compare dataflows to find which of their properties reduces it.
Our contributions are:

\begin{itemize}
\item \textbf{Characterization} (\S\ref{sec:char}).
Fixing placement and the obvious synchronization costs inside \ik gains 1.20--1.73$\times$, and further changes within its operator-at-a-time execution gain nothing.
With node-local weights and two global barriers per layer, threads still wait for 20--48\,\% of each token, mostly for the last thread of their own node; on Emerald Rapids, the share grows with the number of threads per barrier and with fewer bytes per token.

\item \textbf{Dataflow comparison} (\S\ref{sec:dataflow}).
On identical kernels, a row-owned dataflow without cross-node reductions loses to the Megatron-style column split in all 24 configurations, by 5--46\,\%.
The number of global barriers (5 vs.\ 2 per layer) explains a minority of the gap; the traces attribute most of it to late threads and the leaders' serial work, and a controlled variant clears the row-owned dataflow's cross-node writes.

\item \textbf{Evaluation} (\S\ref{sec:eval}).
Where decoding is DRAM-bound, the engine decodes 1.53--2.05$\times$ faster than \ik with all cores and, with Q4\_0, 1.39--2.25$\times$ faster than ArcLight, at \ik's accuracy and with less energy per token.
It keeps this advantage at a context depth of 8192, also against \ik with its KV cache interleaved over the nodes (by default \ik puts the whole cache on one node).
\end{itemize}

\section{Background and Motivation}\label{sec:background}

\paragraph{Ternary LLMs on CPUs}
Ternary models such as BitNet b1.58~\cite{ma2024era} restrict every weight of the linear layers to $\{-1, 0, +1\}$ times a per-tensor or per-row scale and quantize activations to int8.
CPU runtimes for them include llama.cpp~\cite{llamacpp} (TQ1\_0, TQ2\_0), bitnet.cpp~\cite{wang2025bitnetcpp}, the table-lookup kernels of T-MAC~\cite{wei2024tmac} and Vec-LUT~\cite{li2026veclut}, and \ik~\cite{ikllama}, a fork of llama.cpp with its own ternary formats IQ1\_BN and IQ2\_BN.
In our measurements, the 2-bit formats of these runtimes decode at about the same speed on one node, and \ik is the fastest once several nodes are used; it is our baseline.
Decoding multiplies the activation vector with every weight matrix, so each token reads all weights once: on one \icl node with 16 threads, the DRAM reads per token of Llama3-8B are 1.01--1.04 times the weight bytes in every format we measured, and the 2-bit formats reach 72--78\,\% of the node's read bandwidth, so going faster requires more nodes.

\paragraph{Multi-node servers}
A dual-socket server with sub-NUMA clustering exposes each socket as two NUMA nodes with their own memory controllers (Table~\ref{tab:platforms}).
Reading another node's memory costs little within a socket on \icl but 38\,\% across sockets, and 22--23\,\% for any remote node on \emr; interleaving memory over the four nodes yields only 48--55\,\% of the bandwidth available when every thread reads its own node.
Using the whole machine therefore requires that each thread reads weights placed on its own node.

\paragraph{Existing runtimes stop at half the bandwidth}
The llama.cpp family offers two ways to use several nodes: interleaving the weights, and \texttt{--numa distribute}, which binds threads to nodes and lets the first touch of the memory-mapped weights place each page near its first reader.
On \icl with Llama3-8B, interleaving reaches 2.08$\times$ the single-node speed and \texttt{distribute} 2.44$\times$ (174\,GB/s), only 49\,\% of the machine's local bandwidth (48\,\% by DRAM counters in our main batch, \S\ref{sec:eval-bw}); with all weights on one node, 64 threads are no faster than 16.
On \emr, \ik with 104 threads decodes 3--12\,\% slower than with 64 threads on all three models, and a user of llama.cpp has reported a tenfold drop in decode throughput between 16 and 17 threads on another \emr machine~\cite{llamacpp20972}.
Two properties of ggml, the tensor library under llama.cpp and \ik, suggest why.
It executes the model graph one operator at a time, each a parallel region that ends in a global barrier, about 17 times per layer; and operators on a single row (RMS norm, residual add, activation quantization) run on one thread while all others wait, after which threads on all nodes read the result.
Section~\ref{sec:char} measures what these costs amount to once the weights are placed correctly.

\section{Methodology}\label{sec:method}

\begin{table}[t]
\caption{Platforms. Read bandwidths measured with every thread reading memory of its own node, unless noted.}
\label{tab:platforms}
\centering
\small
\setlength{\tabcolsep}{2.5pt}
\resizebox{\columnwidth}{!}{%
\begin{tabular}{@{}lcc@{}}
\toprule
& \icl & \emr \\
\midrule
CPUs (Xeon Platinum) & 2 $\times$ 8358 & 2 $\times$ 8555C \\
NUMA nodes (SNC2) & 4 $\times$ 16 cores & 4 $\times$ 26 cores \\
L3 per socket & 48\,MiB & 260\,MiB \\
Read BW: node / socket & 90.3 / 178.2\,GB/s & 157.5 / 320\,GB/s \\
Read BW: machine & 354.2\,GB/s & 610.4\,GB/s \\
Read BW: interleaved & 170\,GB/s (48\,\%) & 334\,GB/s (55\,\%) \\
int8 peak (machine) & VNNI 44\,Tops & VNNI 73 / AMX 552\,Tops \\
NUMA balancing & off & on \\
\bottomrule
\end{tabular}}
\end{table}

\paragraph{Platforms}
We use two dual-socket Intel servers of consecutive generations with four NUMA nodes each (Table~\ref{tab:platforms}); the \emr machine has 5.4$\times$ the L3 capacity of the \icl machine.
Both run Linux 6.8 and are shared with other users; frequency scaling is left at its default (we record decode-phase frequencies instead, \S\ref{sec:char-counters}), and \emr runs automatic NUMA balancing.

\paragraph{Models}
We use three ternary Llama-architecture models: Llama3-8B-1.58-100B-tokens (HF1BitLLM) with eight KV heads, and Falcon3-10B-Instruct-1.58bit and Falcon3-1B-Instruct-1.58bit (TII) with four.
Each is converted to F16 and quantized to \ik's IQ2\_BN (2 bits per ternary weight, exact) for the main experiments and with llama.cpp to Q2\_K, Q4\_0 and Q4\_K\_M for the format study (\S\ref{sec:char-bytes}); the output projection is Q6\_K in all files.
With IQ2\_BN, a decoded token reads 2.19\,GB (Llama3-8B), 2.73\,GB (Falcon3-10B) and 0.51\,GB (Falcon3-1B) of weights and KV cache, computed from the GGUF headers.

\paragraph{Systems}
\ik is ik\_llama.cpp at commit 0821d62 with one thread per physical core (64 on \icl, 104 on \emr), \texttt{--numa distribute} and memory-mapped weights, which we evict from the page cache and touch once with \ik before each model so that each page lands on the node of its reader.
\ikp is \ik with our four patches (Table~\ref{tab:ladder}), each switchable at run time.
\Row and \Col are the two dataflows of our engine (\S\ref{sec:dataflow}); it pins one thread to each physical core, links \ik's kernels and binds weights, KV cache and per-node buffers to their node with \texttt{mbind}, so automatic NUMA balancing moves practically none of its pages, against $10^5$ or more of \ik's per run.
Later builds only add switches that are off by default.
Below four nodes, \ik and \ikp run with \texttt{--numa numactl} and threads and memory bound to the nodes used.
Since \ik allocates its KV cache on one node (\S\ref{sec:eval-depth}), for long contexts and batches we also run it with the cache interleaved (\texttt{numactl --interleave=all}) and compare against the faster configuration.

\paragraph{Protocol}
We measure the rate of generating 128 tokens (tg128) after a context of a given depth (0 unless stated), with \texttt{llama-bench} for \ik and the engine's own timer for \Row and \Col (after 4 warm-up tokens), and batched decoding with \texttt{llama-batched-bench}.
What a table or figure compares was measured together: each of 5--6 rounds runs every variant once, in rotating order, and we report median speeds and, for ratios, the median of the per-round ratios.
The speed of one binary drifted by up to 10\,\% from day to day while these ratios did not, so we do not compare numbers across batches.
Both machines were shared; a round starts only on an idle machine and is repeated if other processes used more than a few CPUs during it (Appendix~\ref{app:protocol}), and such load, if anything, lowers \Col's speedup over \ik (its per-round value is weakly anticorrelated with the load, Spearman's $\rho$ from $-0.12$ to $-0.46$).
Counters (frequency, UPI traffic, memory-controller reads, RAPL energy) are differences between a 640- and a 128-token run in the same round.

\section{Where Does the Time Go After Placement?}\label{sec:char}

\subsection{Fixing placement within ik}\label{sec:char-place}

\begin{table}[t]
\caption{Decode speed (tg128, depth 0) relative to \ik. Each \ikp step adds one change to the step above. Medians of per-round ratios over 6 interleaved rounds.}
\label{tab:ladder}
\centering
\setlength{\tabcolsep}{3pt}
\resizebox{\columnwidth}{!}{%
\begin{tabular}{@{}lcccccc@{}}
\toprule
& \multicolumn{3}{c}{\icl, 64 threads} & \multicolumn{3}{c}{\emr, 104 threads} \\
\cmidrule(lr){2-4}\cmidrule(l){5-7}
& L-8B & F-10B & F-1B & L-8B & F-10B & F-1B \\
\midrule
\ik (tok/s) & 74.6 & 54.9 & 218.4 & 102.1 & 76.3 & 288.9 \\
+ page migration & 1.03 & 1.03 & 1.14 & 1.03 & 1.00 & 1.06 \\
+ two-level barrier & 1.17 & 1.12 & 1.27 & 1.22 & 1.22 & 1.53 \\
+ split-KV attention & 1.18 & 1.36 & 1.31 & 1.37 & 1.44 & 1.71 \\
+ AVX-512 Q8\_K64 (=\,\ikp) & 1.20 & 1.36 & 1.35 & 1.40 & 1.48 & 1.73 \\
\midrule
\Row (engine, row-owned) & 1.40 & 1.49 & 1.70 & 1.67 & 1.74 & 2.24 \\
\Col (engine, column-split) & \textbf{1.53} & \textbf{1.65} & \textbf{2.00} & \textbf{2.03} & \textbf{2.05} & \textbf{3.23} \\
\Col (tok/s) & 114.4 & 90.5 & 439.8 & 207.2 & 157.8 & 933.2 \\
\bottomrule
\end{tabular}}
\end{table}

We start from the placement \ik already achieves and add the changes of \ikp one at a time (Table~\ref{tab:ladder}); the increments depend on this order, which adds placement first.
After \ik's first pass over its memory-mapped weights, 97\,\% of the weight pages already sit on the node of the thread that reads them, so migrating the rest gains only 0--3\,\% (6--14\,\% on the small Falcon3-1B).
The largest step on both platforms is synchronization: replacing ggml's global barrier by a two-level one (threads meet within their node, then one thread per node meets globally) gains 10--14\,\% on \icl and 20--44\,\% on \emr.
Splitting single-query attention over KV chunks helps models whose few KV heads leave most threads idle in \ik's schedule: the Falcon3 models, and on \emr also Llama3-8B ($8 \times 256 < 32 \times 104$, \ik's threshold for splitting).
Removing one change at a time from \ikp on \emr gives the same order: without the two-level barrier it loses 18--35\,\%, without split attention 9--17\,\%, without migration 1--9\,\% and without the AVX-512 quantizer 0--2\,\%.
Beyond this, local changes inside ggml stop paying off: spreading the single-row operators over all threads slowed the Falcon3 models by 5--9\,\%, and three variants of node-local activation quantization stayed within $\pm 2$\,\%, because the activation vector still has to be gathered from all nodes before the next operator.
What remains is ggml's execution model, one parallel region and one global barrier per operator, about 17 per layer.
Replacing it by our engine (\S\ref{sec:dataflow}), with the same kernels and only a few barriers per layer, gives 1.40--2.24$\times$ (\Row) and 1.53--3.23$\times$ (\Col) over \ik.

\subsection{How much of a token is synchronization?}\label{sec:char-sync}

\begin{figure*}[t]
\centering
\includegraphics[width=\textwidth]{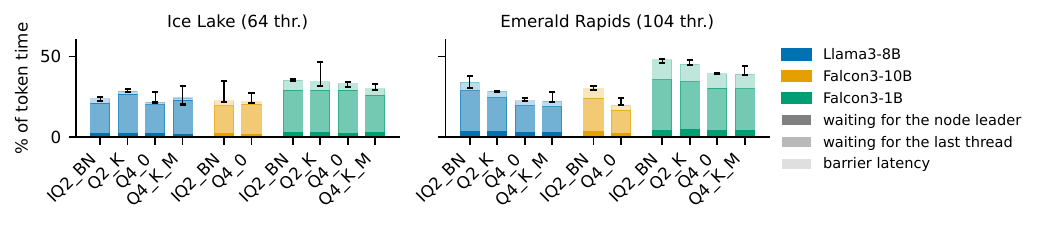}
\vspace{-0.25in}
\caption{Waiting at synchronization points as a share of \Col's per-token decode time (mean over threads; tg128, depth 0; median of 5 interleaved rounds, whiskers: min--max). Split by per-thread barrier traces into waiting for the node leader's serial normalization and quantization (dark), waiting for the last thread to arrive (medium) and the barriers' own latency (light).}
\label{fig:sync}
\end{figure*}

The engine records, per token, how long each thread waits at global and node barriers and for its node leader, which normalizes and quantizes each stage's input alone.
With node-local weights and 2 instead of about 17 global barriers per layer, threads still wait for 22--35\,\% of each token on \icl and 20--48\,\% on \emr (Fig.~\ref{fig:sync}).
Recording each thread's entry and exit at every barrier for 32 tokens, which changes the speed by about 1\,\%, splits this wait: a thread waits for the last thread from its own entry until the last entry, and for the barrier itself from then until its exit.
Waiting for the last thread is 17--26\,\% of the token on \icl and 14--31\,\% on \emr (65--86\,\% of the wait), barrier latency 1--6 and 3--12\,\%, and waiting for the node leader 2--3 and 3--5\,\%.
More than half of the late arrivals are threads of the waiting thread's own node, which read equal shares of node-local weights yet finish at different times; we have not identified the cause.
The share grows with the number of threads per barrier: with IQ2\_BN and 13, 20 and 26 cores per node of \emr, \Col's threads wait for 18--31, 23--42 and 29--48\,\% of the token, mostly through late threads, while other processes kept a median of only 0.4--0.5 CPUs busy at every core count and the share does not correlate with them (Spearman's $\rho$ of $-0.05$).
On \icl the share rises similarly (13--20, 18--27 and 24--34\,\% with 8, 12 and 16 cores per node), but about two foreign CPUs were busy throughout, and once every core is in use they can only run on the sibling hyperthreads of our threads or preempt them, so this trend is partly confounded with interference.
The share is largest for Falcon3-1B, which reads only 0.5\,GB per token; on \emr, where it runs largely from L3 (\S\ref{sec:eval-bw}), nearly half of each token is spent waiting, and with 18 thin layers barrier latency alone is 9--12\,\%.

\subsection{Fewer bytes, more synchronization}\label{sec:char-bytes}

\begin{figure}[t]
\centering
\includegraphics[width=\columnwidth]{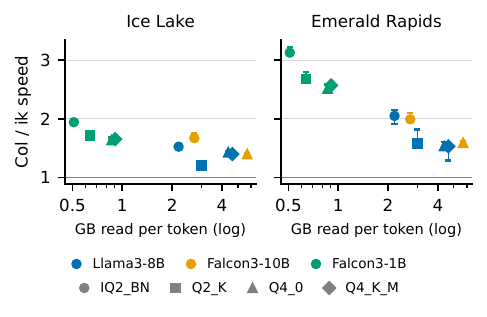}
\vspace{-0.3in}
\caption{Decode speedup of \Col over \ik against the bytes read per token, three models in four formats (tg128, depth 0; median of per-round ratios, 5 rounds).}
\label{fig:bytes}
\end{figure}

If the waiting cost a roughly fixed time per token while the streaming time grew with the bytes read, fewer bytes per token would mean a larger synchronization share and a larger gain from a better dataflow.
We test this by running each model in up to four formats: the ternary IQ2\_BN and the non-ternary Q2\_K, Q4\_0 and Q4\_K\_M, which read 1.3--2.1$\times$ as many bytes.
On \emr the relation holds for both models with four formats: from IQ2\_BN to Q4\_K\_M, the share of \Col falls from 34 to 22\,\% for Llama3-8B and from 48 to 39\,\% for Falcon3-1B, and the speedup over \ik from 2.05 to 1.53$\times$ and from 3.12 to 2.57$\times$ (Fig.~\ref{fig:bytes}).
On \icl it is weak: the shares of Falcon3-1B stay within 30--35\,\%, and Llama3-8B in Q2\_K has the highest share and the lowest speedup (1.20$\times$), an exception that coincides with \ik already using more bandwidth with Q2\_K than with IQ2\_BN there (51 vs.\ 46\,\%).
Across models the ordering holds on both machines: the model with the fewest bytes per token has the largest share and the largest speedup in every format.
Quantitatively, the absolute wait is not fixed: it grows with the work between barriers (for \Col with Llama3-8B on \icl from 1.9\,ms per token with IQ2\_BN to 3.2--3.4\,ms with Q4), because so does the load imbalance a barrier absorbs.

\subsection{Ruling out frequency and placement}\label{sec:char-counters}

Could the engine simply place data better than \ik in ways the ladder misses?
Decode-phase core frequencies agree within 8\,\% (\emr: 2.89\,GHz for all; \icl: 2.98--3.25\,GHz for \ik, 2.93--3.02 for \Col), so the speedup is not from frequency.
Cross-socket traffic (UPI data flits per token) is 4.1--7.3$\times$ higher for \ik than for \Col on \emr and 5.6--7.1$\times$ on \icl although both read the same weight bytes; it comes from each layer's activation exchange and synchronization.
The share of demand loads served by remote DRAM (33--58\,\% for \ik on \icl, 1--2\,\% for \Col) misleads, as the streamed weights arrive almost entirely through the prefetcher (L3-miss stall cycles are 0.5--10\,\%).

\section{Row-Owned vs.\ Column-Split Dataflow}\label{sec:dataflow}

To find which property of a dataflow matters, we built a decode engine that keeps \ik's kernels and GGUF weights and runs two dataflows that differ in how often all threads meet and how much data crosses nodes.

\subsection{The engine}\label{sec:df-engine}

The engine is one process with one thread pinned to each physical core.
At load time each weight matrix is cut into one slice per node and one range of rows per thread, and each slice is copied into memory of its node.
Threads compute their rows with \ik's matrix--vector kernels and quantize activations with \ik's quantizers, so both dataflows perform \ik's arithmetic on the same weights.
Before each matrix stage, one leader thread per node normalizes and quantizes the stage's input into a node-local buffer.
Element-wise operations (RoPE, SiLU, residual add, writing the f16 KV cache) are done by each thread on the rows it just computed, without extra barriers; barriers are two-level, as in \ikp.

\subsection{Two dataflows}\label{sec:df-two}

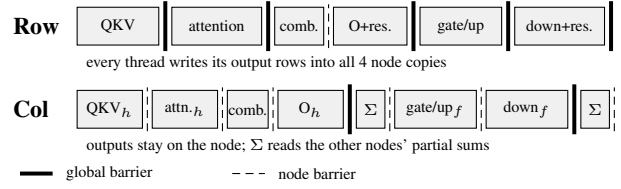
\begin{figure}[t]
\centering
\begin{tikzpicture}[x=1.08mm,y=1mm,font=\tiny,
  st/.style={draw,fill=black!6},
  gb/.style={line width=1.5pt},
  nb/.style={densely dashed,line width=0.4pt}]
\node[anchor=west,font=\footnotesize\bfseries] at (0,13.5) {Row};
\draw[st] (9.0,11) rectangle (19.0,16) node[midway] {QKV};
\draw[gb] (19.8,10.2) -- (19.8,16.8);
\draw[st] (20.6,11) rectangle (31.6,16) node[midway] {attention};
\draw[gb] (32.4,10.2) -- (32.4,16.8);
\draw[st] (33.2,11) rectangle (39.2,16) node[midway] {comb.};
\draw[nb] (39.8,10.6) -- (39.8,16.4);
\draw[st] (40.4,11) rectangle (49.4,16) node[midway] {O+res.};
\draw[gb] (50.2,10.2) -- (50.2,16.8);
\draw[st] (51.0,11) rectangle (61.0,16) node[midway] {gate/up};
\draw[gb] (61.8,10.2) -- (61.8,16.8);
\draw[st] (62.6,11) rectangle (73.6,16) node[midway] {down+res.};
\draw[gb] (74.4,10.2) -- (74.4,16.8);
\node[anchor=west,font=\tiny] at (9,8.6) {every thread writes its output rows into all 4 node copies};
\node[anchor=west,font=\footnotesize\bfseries] at (0,2.5) {Col};
\draw[st] (9.0,0) rectangle (17.0,5) node[midway] {QKV$_h$};
\draw[nb] (17.6,-0.4) -- (17.6,5.4);
\draw[st] (18.2,0) rectangle (26.2,5) node[midway] {attn.$_h$};
\draw[nb] (26.8,-0.4) -- (26.8,5.4);
\draw[st] (27.4,0) rectangle (32.4,5) node[midway] {comb.};
\draw[nb] (33.0,-0.4) -- (33.0,5.4);
\draw[st] (33.6,0) rectangle (41.6,5) node[midway] {O$_h$};
\draw[gb] (42.4,-0.8) -- (42.4,5.8);
\draw[st] (43.2,0) rectangle (46.7,5) node[midway] {$\Sigma$};
\draw[nb] (47.3,-0.4) -- (47.3,5.4);
\draw[st] (47.9,0) rectangle (57.9,5) node[midway] {gate/up$_f$};
\draw[nb] (58.5,-0.4) -- (58.5,5.4);
\draw[st] (59.1,0) rectangle (69.1,5) node[midway] {down$_f$};
\draw[gb] (69.9,-0.8) -- (69.9,5.8);
\draw[st] (70.7,0) rectangle (74.2,5) node[midway] {$\Sigma$};
\draw[nb] (74.8,-0.4) -- (74.8,5.4);
\node[anchor=west,font=\tiny] at (9,-2.4) {outputs stay on the node; $\Sigma$ reads the other nodes' partial sums};
\draw[gb] (2,-6) -- (6,-6); \node[anchor=west,font=\tiny] at (6.5,-6) {global barrier};
\draw[nb] (28,-6) -- (32,-6); \node[anchor=west,font=\tiny] at (32.5,-6) {node barrier};
\end{tikzpicture}
\caption{One layer in the two dataflows. \Row splits every matrix by output rows (5 global barriers per layer). \Col gives each node a quarter of the heads ($h$) with their KV cache and of the FFN ($f$); O and down yield partial sums that every node adds up ($\Sigma$) after each of 2 global barriers. Each matrix stage starts with the node leader's norm and quantization (not shown).}
\label{fig:dataflow}
\end{figure}

\paragraph{Row-owned (\Row)}
Every matrix is split by output rows across all threads of all nodes.
Since the next stage's node leaders need whole vectors, each thread writes its output rows into a copy of the vector on every node, a parallel scatter instead of a serial gather.
A layer has five stages (QKV, attention, O with residual, gate/up with SiLU, down with residual), each ending in a global barrier, and no reductions (Fig.~\ref{fig:dataflow}).
Avoiding cross-node reductions is meant to keep the traffic between nodes small; the scatter writes instead generate more cross-socket traffic, although \S\ref{sec:df-barrier} shows that this traffic is not what slows \Row.

\paragraph{Column-split (\Col)}
This is the Megatron-LM decomposition~\cite{shoeybi2019megatron}, which ArcLight~\cite{xu2026arclight} applies to the NUMA nodes of a CPU.
Node $n$ owns a quarter of the query heads with their KV heads and KV cache, and a quarter of the FFN's intermediate dimension, and computes them with node-level barriers only.
O and down are split by input columns, so each node obtains a partial sum of the whole hidden vector; after a global barrier, every node adds the four partial sums into its residual copy in a fixed order, so the copies stay identical.
A layer has two global and six node barriers, at the price of two reductions in which each node reads three remote vectors of the hidden size (16\,KiB for Llama3-8B).

\subsection{Comparison}\label{sec:df-cmp}

\begin{table}[t]
\caption{Decode speed of \Col relative to \Row (median of per-round ratios over 5 interleaved rounds; tg128 starting at depth 0 or 4096).}
\label{tab:tpcol}
\centering
\setlength{\tabcolsep}{4pt}
\resizebox{\columnwidth}{!}{%
\begin{tabular}{@{}llcccc@{}}
\toprule
& & \multicolumn{2}{c}{IQ2\_BN} & \multicolumn{2}{c}{Q4\_0} \\
\cmidrule(lr){3-4}\cmidrule(l){5-6}
& Model & depth 0 & 4096 & depth 0 & 4096 \\
\midrule
\multirow{3}{*}{\icl} & Llama3-8B & 1.09 & 1.06 & 1.05 & 1.06 \\
& Falcon3-10B & 1.11 & 1.11 & 1.11 & 1.08 \\
& Falcon3-1B & 1.21 & 1.19 & 1.15 & 1.16 \\
\midrule
\multirow{3}{*}{\emr} & Llama3-8B & 1.21 & 1.18 & 1.14 & 1.11 \\
& Falcon3-10B & 1.20 & 1.24 & 1.11 & 1.20 \\
& Falcon3-1B & 1.46 & 1.41 & 1.32 & 1.31 \\
\bottomrule
\end{tabular}}
\end{table}

\Col is faster than \Row in all 24 configurations of Table~\ref{tab:tpcol}, by 5--21\,\% on \icl and 11--46\,\% on \emr.
The gain is largest for Falcon3-1B, larger with IQ2\_BN than with the 1.7--2.1$\times$ larger Q4\_0 in 11 of 12 cases, about twice as large on \emr, and persists at depth 4096, in line with the wait shares of Section~\ref{sec:char}.

\begin{figure*}[t]
\centering
\includegraphics[width=\textwidth]{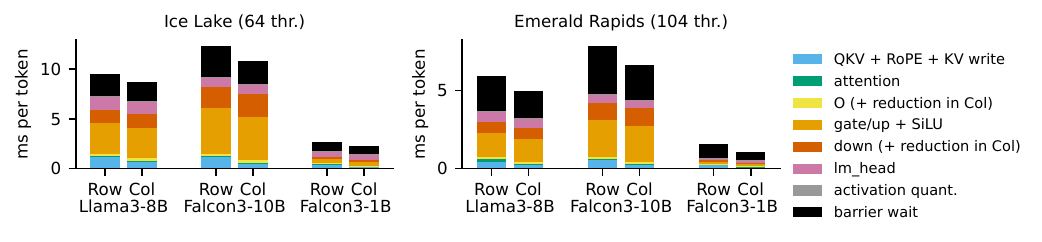}
\vspace{-0.25in}
\caption{Per-token decode time of the engine by stage (mean over threads; IQ2\_BN, tg128, depth 0; median of 5 rounds). In \Col, O and down include the reduction of the partial sums.}
\label{fig:stages}
\end{figure*}

Two stages shrink (Fig.~\ref{fig:stages}).
The barrier wait falls by 7--23\,\% on \icl and 25--35\,\% on \emr.
The QKV stage, whose time excludes the barrier wait, becomes 36--57\,\% shorter: in \Row, K and V rows go to the node that holds the KV block of the token's position and Q to a shared buffer, so most threads write across nodes, whereas in \Col a node's heads, Q and KV cache stay on the node.
The reductions add at most 0.22\,ms per token to O and down, while the token time drops by 0.38--1.45\,ms: the partial sums never travel as messages, each node reads them from shared memory after the barrier.
\Col also gives the node leaders less serial work: the inputs of O and down are the node's own slices of the attention output and the FFN activation, so a leader quantizes 39--48\,\% as many elements per layer as in \Row, and on \emr the slowest leader's time per token falls by 37--52\,\%.

\subsection{Isolating the number of barriers}\label{sec:df-barrier}

\begin{table}[t]
\caption{Barrier isolation on \emr (tg128, depth 0, 5 interleaved rounds). Gap: \Row minus \Col token time. Share of the gap recovered when \Col gets 3 more global barriers per layer, by promoting 3 of its node barriers (the same 5 global barriers per layer as \Row, same work and data) or by adding 3 empty ones; and the latency of one empty global barrier among 104 threads.}
\label{tab:barrier}
\centering
\small
\setlength{\tabcolsep}{3pt}
\resizebox{\columnwidth}{!}{%
\begin{tabular}{@{}llcccc@{}}
\toprule
Model & Format & Gap (ms) & +3 promoted & +3 empty & $\mu$s / barrier \\
\midrule
Llama3-8B & IQ2\_BN & 1.01 & 9\,\% & 26\,\% & 2.7 \\
 & Q4\_0 & 1.35 & 5\,\% & 23\,\% & 3.3 \\
Falcon3-10B & IQ2\_BN & 1.38 & 26\,\% & 38\,\% & 4.4 \\
 & Q4\_0 & 1.05 & 14\,\% & 16\,\% & 1.4 \\
Falcon3-1B & IQ2\_BN & 0.50 & 15\,\% & 32\,\% & 3.0 \\
 & Q4\_0 & 0.60 & 5\,\% & 23\,\% & 2.5 \\
\bottomrule
\end{tabular}}
\end{table}

The two dataflows differ both in the number of global barriers and in where they read and write, so we separate the two in the engine (Table~\ref{tab:barrier}).
Promoting three of \Col's node barriers to global ones gives it the five global barriers per layer of \Row with its own data placement, and adding empty global barriers right after an existing one, when all threads have just met, measures barrier latency without imbalance; neither changes the computed values.

The number of barriers explains only a minority of the gap: five global barriers per layer slow \Col by 5--26\,\% of the \Row--\Col difference, and three complete extra barriers by 16--38\,\%.
An empty global barrier among 104 threads costs 1.4--4.4\,$\mu$s, so \Col's two per layer take 1--10\,\% of its token time, a small part of its 20--48\,\% wait.
On \icl the round-to-round variation of the shared machine (4--11\,\%) exceeded these effects (Appendix~\ref{app:barrier-icl}).

\Row also writes across nodes (each thread writes its output rows into every node's copy and its K and V rows to the node holding the token's KV block), with 4.1--5.0$\times$ the cross-socket traffic of \Col on \emr and 1.8--2.4$\times$ on \icl; its leaders quantize more than twice as many elements; and it splits attention by position rather than by head.
In the traced batches (not that of Table~\ref{tab:barrier}), the gap of 0.43--2.67\,ms per token consists of \Row's longer waits for late threads (up to 1.29\,ms) and for the leaders (up to 0.54\,ms) and its longer QKV, attention and gate/up stages (up to 0.62, 0.15 and 0.19\,ms), less \Col's reductions (up to 0.23\,ms); barrier latency adds at most 0.08\,ms, and the residual is below 0.14\,ms (Appendix~\ref{app:gap}).
The cross-node writes of the stage outputs do not explain the extra waiting: a variant of \Row that writes the outputs of O, gate/up and down only to its own node and copies the other nodes' rows after the barrier (bit-identical results) is 1--7\,\% slower and waits as long for late threads ($-4$ to $+3$\,\%).
\Row's extra late arrivals occur mainly at the barriers after QKV and attention, which also differ in the K and V writes and the attention split.

\subsection{Comparison with ArcLight}\label{sec:df-arclight}

\begin{table}[t]
\caption{Decode speed with Q4\_0 weights (tok/s, tg128, depth 0; median of 5 interleaved rounds). ArcLight built for AVX2; a build with \texttt{-march=native} was within 3\,\%. Last row: median of the per-round ratios, which can differ from the ratio of the medians by a few percent. Configuration in Appendix~\ref{app:runtimes}.}
\label{tab:arclight}
\centering
\setlength{\tabcolsep}{2.6pt}
\resizebox{\columnwidth}{!}{%
\begin{tabular}{@{}lcccccc@{}}
\toprule
& \multicolumn{3}{c}{\icl, 64 threads} & \multicolumn{3}{c}{\emr, 104 threads} \\
\cmidrule(lr){2-4}\cmidrule(l){5-7}
& L-8B & F-10B & F-1B & L-8B & F-10B & F-1B \\
\midrule
llama.cpp (\texttt{distribute}) & 17.0 & 12.8 & 84.9 & 60.6 & 48.3 & 195.0 \\
\ik & 41.6 & 32.7 & 162.9 & 68.4 & 48.9 & 225.1 \\
ArcLight~\cite{xu2026arclight} & 38.9 & 32.8 & 119.6 & 56.3 & 45.5 & 162.3 \\
\ikp & 44.7 & 38.3 & 193.1 & 80.3 & 60.9 & 319.0 \\
\Row & 55.7 & 40.5 & 234.6 & 88.7 & 72.9 & 392.3 \\
\Col & \textbf{59.4} & \textbf{45.6} & \textbf{270.5} & \textbf{101.6} & \textbf{77.8} & \textbf{528.4} \\
\midrule
\Col / ArcLight & 1.53 & 1.39 & 2.25 & 1.75 & 1.74 & 3.23 \\
\bottomrule
\end{tabular}}
\end{table}

ArcLight implements the same decomposition on top of llama.cpp's operators and kernels and supports only F16, Q4\_0, Q8\_0 and Q6\_K weights, so we compare with Q4\_0, the format it was evaluated with~\cite{xu2026arclight} (Table~\ref{tab:arclight}).
ArcLight runs at 0.72--1.00$\times$ the speed of \ik, and \Col is 1.39--2.25$\times$ (\icl) and 1.74--3.23$\times$ (\emr) faster than ArcLight (llama.cpp is unusually slow on \icl because it repacks Q4\_0 weights into a buffer on one node).
On one node of \emr, ArcLight is within 4--11\,\% of \ik and \Col only 1.05--1.23$\times$ faster, whereas on four nodes \Col is 1.81--3.59$\times$ faster in the same batch, so the gap arises across nodes.
The decomposition alone therefore does not deliver the speedup; across nodes, ArcLight executes llama.cpp's operator graph with node-level barriers inside its tensor-parallel subgraphs and global ones at their boundaries~\cite{xu2026arclight}, whereas \Col runs each stage as one parallel region.

\section{Evaluation}\label{sec:eval}

\subsection{Decode throughput}\label{sec:eval-tput}

At depth 0, \Col decodes 1.53, 1.65 and 2.00$\times$ as fast as \ik on \icl and 2.03, 2.05 and 3.23$\times$ on \emr for Llama3-8B, Falcon3-10B and Falcon3-1B (Table~\ref{tab:ladder}), with 95\,\% bootstrap intervals within 4\,\% of the medians; on another \icl machine and other days, the ratios of the engine changed by less than 5\,\%.
\ik is not always fastest with all cores: on \emr it is 6--14\,\% faster with 64 or 80 of its 104 cores, and on \icl Falcon3-10B is 6\,\% faster with 32 of 64.
Against \ik at its fastest thread count, \Col is 1.53, 1.54 and 1.93$\times$ faster on \icl and 1.89, 1.92 and 2.78$\times$ on \emr.
Among existing runtimes \ik is the fastest on both machines: on \emr, llama.cpp (TQ2\_0, Q4\_0) and bitnet.cpp (I2\_S) reach 0.59--0.85$\times$ of its speed, and the two weight-replication pull requests~\cite{llamacpp27986,ikllama2396} do not improve on it (Appendix~\ref{app:runtimes}).

\subsection{Bandwidth utilization}\label{sec:eval-bw}

\begin{figure*}[t]
\centering
\includegraphics[width=\textwidth]{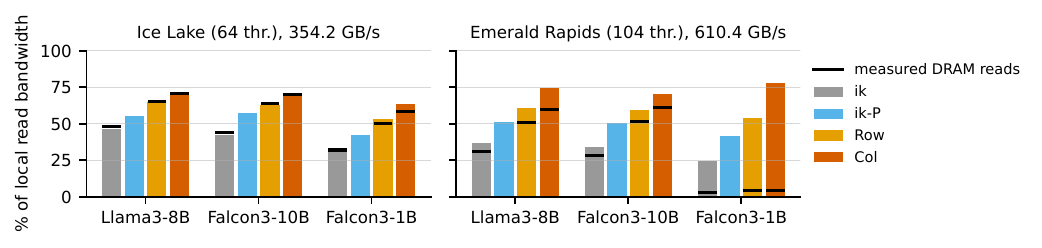}
\vspace{-0.25in}
\caption{Effective read bandwidth as a share of the machine's local read bandwidth (tg128, depth 0, IQ2\_BN). Bars: speed $\times$ weight and KV bytes per token. Markers: speed $\times$ DRAM reads per token from memory-controller counters (separate batch, 3 rounds; not measured for \ikp).}
\label{fig:roofline}
\end{figure*}

Decode speed times bytes per token gives a kernel-independent effective bandwidth (Fig.~\ref{fig:roofline}).
\Col reaches 63--71\,\% of the machine's local read bandwidth on \icl and 70--78\,\% on \emr, against 31--46\,\% and 24--37\,\% for \ik.
Memory-controller counters confirm this on \icl, whose 48\,MiB L3 per socket holds little: DRAM reads per token are 92--103\,\% of the weight and KV bytes, and \Col uses 59--71\,\% of the bandwidth against 32--48\,\% for \ik.
On \emr, 9--20\,\% of the bytes of the two larger models come from the 260\,MiB L3 per socket, so by measured DRAM reads \Col uses 60--61\,\% and \ik 28--31\,\%.
Falcon3-1B (482\,MiB of weights) runs on \emr almost entirely from L3, with only 0.03--0.06 of its 0.53\,GB per token from DRAM; its 3.2$\times$ speedup there measures on-chip synchronization and computation, consistent with its 39--48\,\% barrier share (Fig.~\ref{fig:sync}).
On \icl, this is close to what every implementation reaches on a single node (\S\ref{sec:eval-scale}), so the remaining limit there is the kernel.

\subsection{Scaling over nodes}\label{sec:eval-scale}

\begin{figure*}[t]
\centering
\includegraphics[width=\textwidth]{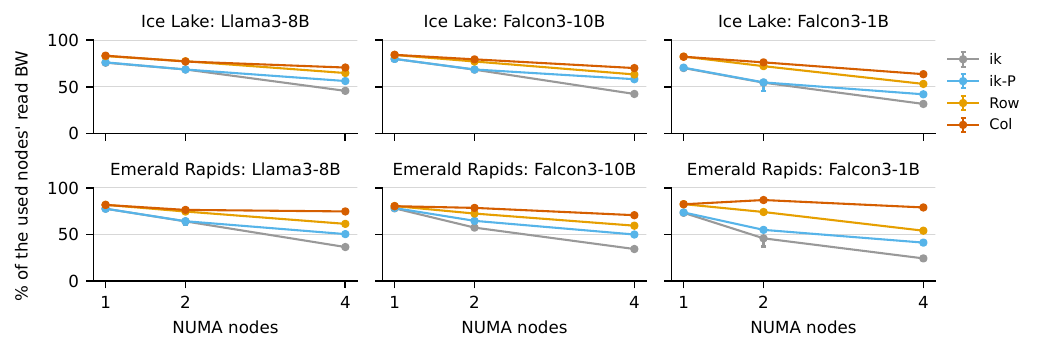}
\vspace{-0.25in}
\caption{Effective read bandwidth as a share of the read bandwidth of the nodes used (Table~\ref{tab:platforms}), tg128, depth 0, IQ2\_BN; a separate batch from Fig.~\ref{fig:roofline}. Below 4 nodes, \ikp's migration and two-level barrier do not apply.}
\label{fig:scale}
\end{figure*}

On one node the implementations differ by at most 17\,\% and all reach 70--84\,\% of the node's bandwidth (Fig.~\ref{fig:scale}).
From one node to four, \ik speeds up only 1.77--2.36$\times$ (\icl) and 1.29--1.82$\times$ (\emr), while \Col speeds up 3.0--3.3$\times$ and 3.4--3.7$\times$ (on \emr partly helped by the larger L3 of more nodes).
The engine's advantage therefore arises across nodes.

\subsection{Context depth and KV-cache placement}\label{sec:eval-depth}

\begin{figure*}[t]
\centering
\includegraphics[width=\textwidth]{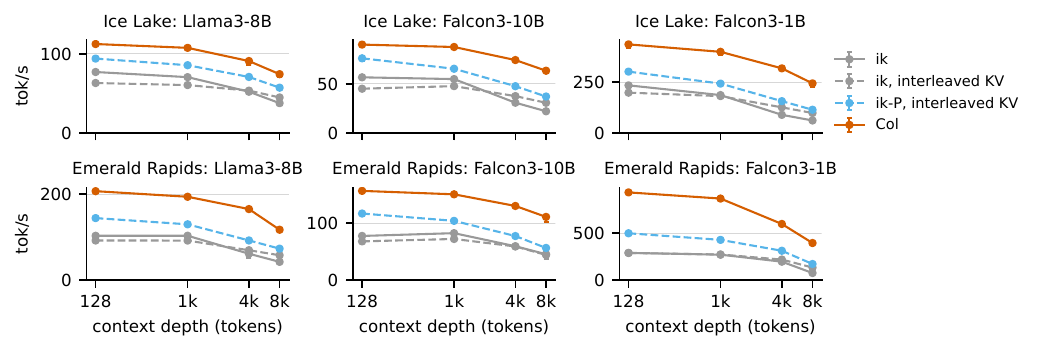}
\vspace{-0.25in}
\caption{Decode speed (tg128) after a context of the given depth, 5 interleaved rounds. \ik with its default KV cache (on one node) and with the cache interleaved over all nodes; ratios in the text are against the faster of the two.}
\label{fig:depth}
\end{figure*}

Long contexts exposed a placement issue in the baseline: \ik allocates the whole KV cache on a single node.
Interleaving it with \texttt{numactl --interleave=all}, which needs no code change, makes \ik faster at depth 8192 by 19, 38 and 60\,\% on \icl and by 36, $-3$ and 76\,\% on \emr (Llama3-8B, Falcon3-10B, Falcon3-1B), but slower at depth 128 by 15--20\,\% on \icl and 11--13\,\% on \emr for the larger models (Appendix~\ref{app:open}).
On \emr, automatic NUMA balancing moves a varying part of the cache during each run (the most loaded node held 27--55\,\% of it), so the default placement there is also unstable (Appendix~\ref{app:kv}).
Against the faster of the two \ik configurations at each depth (Fig.~\ref{fig:depth}), \Col is 1.46, 1.59 and 1.86$\times$ faster on \icl at depth 128 and 1.66, 2.08 and 2.49$\times$ at 8192, and on \emr 2.00, 2.02 and 3.23$\times$ at 128 and 2.02, 2.44 and 2.93$\times$ at 8192: its advantage holds at every depth, growing by 13--34\,\% on \icl but not on \emr.
Against \ik's default placement alone, the depth-8192 ratios would have been overstated as 1.97--3.94$\times$ and 2.44--5.17$\times$.
With the interleaved cache, \ikp is 1.17--1.33$\times$ (\icl) and 1.24--1.71$\times$ (\emr) faster than the faster \ik at every depth.

\subsection{Batches, prefill, energy and accuracy}\label{sec:eval-batch}

\paragraph{Batches}
The engine decodes a single sequence; with \texttt{llama-batched-bench}, \ikp's gain over \ik is 1.21--1.31$\times$ (\icl) and 1.38--1.59$\times$ (\emr) at one sequence but only 1.02--1.17$\times$ at eight or more, while batching itself raises \ik's throughput 1.9--3.4$\times$ (Appendix~\ref{app:batch}).
\Col therefore targets the latency of a single sequence, as in interactive decoding.

\paragraph{Prefill}
The engine does not process prompts; prefill is a compute-bound matrix product, which we accelerate separately inside \ik with AMX tiles (Appendix~\ref{app:amx}).
This is 1.43--2.07$\times$ faster than \ik's fastest prefill on one node of \emr but only 1.08--1.39$\times$ on the whole machine, where \ik's per-operator synchronization again limits scaling.

\paragraph{Energy}\label{sec:eval-energy}
Measured with RAPL over the whole machine (package and DRAM), \Col uses 29--45\,\% less energy per token than \ik on \icl and 39--64\,\% less on \emr (Appendix~\ref{app:energy}).
On \icl all implementations run at the package power limit, so the saving follows the speedup; on \emr \Col draws 16--29\,\% more package power than \ik, so the saving is smaller than the speedup.
RAPL covers the whole machine, so the readings include other users' processes; the difference between two run lengths removes only their constant part.

\paragraph{Accuracy}\label{sec:eval-acc}
All paths quantize activations to int8 and differ slightly from an F16 reference with the same ternary weights; on 5100 WikiText-2 tokens, \ik's two decode paths (with and without flash attention) have perplexities within 0.4\,\% of the reference and KL divergences of 0.0030--0.0045.
\Row falls within this spread, and \Col, which quantizes the inputs of O and down per node and thus with finer scales, has the lowest KL divergence of all paths (0.0020--0.0034).
Over 256 greedy tokens, \Col reproduces \ik's output for 7--11 of 32 prompts, as often as \ik's two decode paths agree with each other (6--8); Appendix~\ref{app:acc} gives the details.

\section{Discussion and Limitations}\label{sec:discussion}

\paragraph{Generality}
The effect is not specific to ternary weights: with Q4\_0 and Q4\_K\_M, \Col is still 1.40--2.57$\times$ faster than \ik and threads wait for 20--39\,\% of a token (Figs.~\ref{fig:sync}, \ref{fig:bytes}); formats with even fewer bytes, such as BITCOS at $2 - z$ bits per weight for a zero fraction $z$~\cite{georganas2026bitcos}, should make the dataflow matter more.

\paragraph{Implications for CPU runtimes}
First, on multi-node machines one global barrier per operator, with single-row operators on one thread, costs more than misplaced weights; grouping a layer into a few stages is the larger lever, and beyond that, how the work is divided among threads matters more than the barrier count.
Second, most of the remaining wait is imbalance among a node's threads, which fixed row ranges cannot absorb; assigning rows dynamically within a node is the direct remedy to test.
Spreading the leader's serial work (3--5\,\% of the token) over the node's threads made \Col 3--22\,\% slower, because the threads must then exchange statistics and meet once more per stage; letting every thread quantize the whole vector itself would avoid both.
Third, the KV cache should be placed like the weights: an interleave makes \ik up to 76\,\% faster at depth 8192 than its single-node default.
Fourth, thresholds tuned for one machine break on another: \ik's rule for splitting single-query attention leaves most threads idle at short context for models with four KV heads and, at 104 threads, also for Llama3-8B.

\paragraph{Scope}
\Col supports Llama-architecture models whose head, KV-head and FFN counts are divisible by the number of nodes and whose head size is a multiple of 128, and decodes a single sequence without prefill or batching; it targets latency-bound single-user settings (\S\ref{sec:eval-batch}).
Of the public ternary models, Llama3-8B-1.58 is fine-tuned from a full-precision model, and BitNet b1.58 2B4T, trained ternary, uses squared ReLU and extra sub-layer normalization, which the engine does not implement.
Hybrid-attention models, more than four NUMA nodes, AMD systems and Intel systems with SNC disabled are outside our measurements, and \emr's large L3 makes its Falcon3-1B results reflect on-chip rather than memory behavior.

\paragraph{Threats to validity}
Sharing the machines matters little for the comparisons between implementations, whose variants alternate within every round and are hit alike by other users' load (\S\ref{sec:method}).
Absolute wait shares may be affected: in the trace batches of Fig.~\ref{fig:sync} the share grows with interference (Spearman's $\rho$ of 0.36 on \emr and 0.52 on \icl); on \emr, medians over rounds with at most one busy foreign CPU differ by at most 3.3 points, but \icl never had fewer than 1.8 busy foreign CPUs, so its shares may be a few points high, and its thread-count trend is partly confounded with interference (\S\ref{sec:char-sync}).

\section{Related Work}\label{sec:related}

Numbers quoted here are those reported by the cited works on their platforms.

\paragraph{Multi-node CPU inference}
ArcLight~\cite{xu2026arclight} is the closest prior work: it groups threads by NUMA node, partitions weights in the Megatron-LM style~\cite{shoeybi2019megatron}, synchronizes globally only at the boundaries of its tensor-parallel subgraphs, and reports up to 46\,\% higher throughput than mainstream frameworks.
\Col uses the same decomposition; our contribution is the measurement of what limits decoding once it is in place and the comparison with a row-owned dataflow on identical kernels (on our machines \Col is 1.39--3.23$\times$ faster than ArcLight).
SGLang's CPU backend maps tensor parallelism to NUMA nodes with shared-memory collectives and reports that communication takes 3\,\% of the end-to-end time overall, in deployments centered on large MoE models such as DeepSeek-R1~\cite{sglang2025xeon}, consistent with synchronization weighing less when more bytes are read.
xFasterTransformer distributes inference over sockets and machines~\cite{he2024inference}, KTransformers splits MoE expert weights across sockets in a CPU/GPU hybrid~\cite{chen2025ktransformers}, and two pull requests replicate weights on every node~\cite{llamacpp27986,ikllama2396}.

\paragraph{Synchronization when data movement shrinks}
On CPUs whose 3D-stacked last-level caches hold the weights, \citet{zhang2026cacheresident} find that operator-boundary synchronization becomes a visible bottleneck and relax it to sub-operator dependencies; on GPUs, \citet{chen2026membound} finds that batch-1 decoding reaches a smaller fraction of peak bandwidth the faster the memory, and attributes the gap to kernel-launch overhead.
We show the same shift for DRAM-bound decoding on standard multi-node servers, quantify it against the bytes per token and the threads per barrier, and compare dataflows on identical kernels; our \emr results for Falcon3-1B fall into the cache-resident regime of \citet{zhang2026cacheresident}.

\paragraph{Reducing synchronization on GPUs}
GPU decoders fuse a forward pass into one persistent kernel with few synchronization points per layer, hand-written~\cite{hazy2025nobubbles,hazy2025tp}, generated by agents or search~\cite{li2026forgemegakernel,dong2026adamk,jaber2026amk}, or for multi-die GPUs whose dies resemble NUMA nodes~\cite{chowdhary2026fleet}.
On CPU nodes, we find that reducing the number of synchronization points stops paying off once they are few (\S\ref{sec:df-barrier}).

\paragraph{Ternary and low-bit kernels; AMX}
Ternary CPU work mostly optimizes the kernel on one node or socket: table lookups in T-MAC~\cite{wei2024tmac} and Vec-LUT~\cite{li2026veclut}, lossless kernels in bitnet.cpp~\cite{wang2025bitnetcpp}, Intel's near-roofline 2-bit microkernels~\cite{georganas2025pushing}, the sub-two-bit BITCOS format~\cite{georganas2026bitcos}, and FairyFuse's fused AVX-512 loop on one Xeon~\cite{zuo2026fairyfuse}.
They are complementary: a faster kernel or a smaller format shortens the work between barriers and makes the dataflow matter more.
AMX is used for MoE experts in KTransformers~\cite{chen2025ktransformers}, for sparse compressed models in SparAMX~\cite{abouelhamayed2025sparamx}, and for ternary prefill in Intel's TPP-based framework.
Appendix~\ref{app:amx} reports an AMX ternary prefill inside \ik as a side result.

\section{Conclusion}\label{sec:conclusion}

Placing every weight on the node of its reader is necessary for decoding ternary LLMs on multi-node CPU servers, but threads then still wait for 20--48\,\% of each token, mostly for the slowest threads of their own node.
On the same kernels, a column-split dataflow decodes 1.5--2.1$\times$ faster than the fastest existing ternary runtime where decoding is DRAM-bound, mainly because its threads wait less for each other and for their node leaders.
Balancing the work among a node's threads is the next target, along with integrating such a dataflow into a runtime with batching and prefill and testing it on more nodes and on other vendors' CPUs.

\bibliography{refs}
\bibliographystyle{mlsys2025}

\appendix

\section{Measurement protocol}\label{app:protocol}

Both machines were shared with other users.
A round of a batch started only after the machine had been idle for 20\,s (at most two busy CPUs).
After each run we computed how many CPUs other processes had kept busy on average, as the machine's busy time minus that of the run's own processes, per second; a round in which this exceeded 2.5 CPUs on \icl (3.5 in the later batches) or 4 on \emr in any run was discarded and repeated after a further idle period.
Four early batches had only the idle check before each round: the \icl rows of the paper's \Row-versus-\Col and ArcLight tables, \icl batched decoding (Appendix~\ref{app:batch}) and the other runtimes on \emr (Appendix~\ref{app:runtimes}).
On \emr, medians over the rounds with at most one busy foreign CPU differ from those over all rounds by at most 3.7\,\%.

\section{Accuracy}\label{app:acc}

\begin{table}[h]
\caption{Perplexity relative to the F16 reference (same ternary weights, F16 activations) and mean KL divergence from it, on 5100 WikiText-2 tokens (\icl, 32 threads). ``\ik, no FA'' is \ik's decode path without flash attention, a second correct implementation. \ik and \ikp rows measured 2026-10-06, \Row and \Col (the engine configuration of the paper) 2026-10-16, against the same reference.}
\label{tab:acc}
\centering
\small
\setlength{\tabcolsep}{3pt}
\resizebox{\columnwidth}{!}{%
\begin{tabular}{@{}lcccccc@{}}
\toprule
& \multicolumn{2}{c}{Llama3-8B} & \multicolumn{2}{c}{Falcon3-10B} & \multicolumn{2}{c}{Falcon3-1B} \\
\cmidrule(lr){2-3}\cmidrule(lr){4-5}\cmidrule(l){6-7}
& PPL ratio & KLD & PPL ratio & KLD & PPL ratio & KLD \\
\midrule
\ik (decode path) & 1.0030 & 0.00451 & 0.9993 & 0.00379 & 1.0038 & 0.00297 \\
\ik, no FA & 1.0040 & 0.00448 & 0.9983 & 0.00373 & 1.0030 & 0.00298 \\
\ikp & 1.0030 & 0.00451 & 0.9993 & 0.00379 & 1.0038 & 0.00297 \\
\Row & 1.0035 & 0.00440 & 0.9996 & 0.00377 & 1.0015 & 0.00298 \\
\Col & 1.0023 & 0.00336 & 0.9997 & 0.00200 & 1.0018 & 0.00196 \\
\bottomrule
\end{tabular}}
\end{table}

\paragraph{Greedy generation}
For 32 prompts and 256 greedily generated tokens each, \Col produces the same 256 tokens as \ik's decode path for 11, 8 and 7 prompts (Llama3-8B, Falcon3-10B, Falcon3-1B), with the first difference at a median position of 20, 18 and 23; \Row for 9, 8 and 7 prompts (median positions 26, 30, 34).
\ik without flash attention agrees with \ik for 7, 6 and 8 prompts, diverging at median positions 20, 14 and 24.
\ikp is bit-identical to \ik at this thread count.
Greedy decoding flips near-ties under any change of summation order, so these numbers measure closeness to \ik, not generation quality.

\paragraph{Downstream tasks (\amxp)}
On \emr, \amxp (Appendix~\ref{app:amx}) and \ik score within 1.0 point of each other on HellaSwag (1000 questions), WinoGrande (1267), ARC-Easy (570) and ARC-Challenge (299) for all three models, within one standard error and without a consistent direction (e.g., Llama3-8B 66.6 vs.\ 66.5, 60.06 vs.\ 59.91, 67.54 vs.\ 67.19, 40.47 vs.\ 40.47).

\paragraph{\Col}
\Col, which quantizes the inputs of O and down per node and thus with finer scales than one scale per row, has the lowest KL divergence from the reference of all paths (Table~\ref{tab:acc}; top-token agreement 96.4, 97.3 and 96.9\,\%).
On one node its perplexity is bit-identical to \Row's.
The \Row rows are the current engine configuration; an earlier version with a different attention schedule, measured on 2026-10-06, gave 0.00446, 0.00374 and 0.00298.

\section{Energy}\label{app:energy}

\begin{table}[h]
\caption{Whole-machine energy per decoded token (package + DRAM, J/token, RAPL), tokens 128--640, median of 5 interleaved rounds, and package power while decoding.}
\label{tab:energy}
\centering
\small
\setlength{\tabcolsep}{3pt}
\resizebox{\columnwidth}{!}{%
\begin{tabular}{@{}lcccccc@{}}
\toprule
& \multicolumn{3}{c}{\icl} & \multicolumn{3}{c}{\emr} \\
\cmidrule(lr){2-4}\cmidrule(l){5-7}
& L-8B & F-10B & F-1B & L-8B & F-10B & F-1B \\
\midrule
\ik & 8.12 & 10.92 & 2.73 & 4.25 & 6.36 & 1.40 \\
\ikp & 7.34 & 9.12 & 2.30 & 3.56 & 4.52 & 0.86 \\
\Row & 6.12 & 7.89 & 1.77 & 3.07 & 3.90 & 0.70 \\
\Col & \textbf{5.74} & \textbf{7.23} & \textbf{1.49} & \textbf{2.60} & \textbf{3.38} & \textbf{0.50} \\
\midrule
\Col vs.\ \ik & $-29$\,\% & $-34$\,\% & $-45$\,\% & $-39$\,\% & $-47$\,\% & $-64$\,\% \\
Package W, \ik / \Col & 498 / 504 & 498 / 509 & 506 / 533 & 360 / 417 & 352 / 420 & 332 / 429 \\
\bottomrule
\end{tabular}}
\end{table}

The readings cover both sockets and all DRAM, including idle cores and other users' processes; subtracting the 128-token run from the 640-token run removes only their constant part.

\section{Other runtimes}\label{app:runtimes}

\begin{table}[h]
\caption{Decode speed on \emr (104 threads, tg128, depth 0) relative to \ik. ``mirror'': the pull requests that replicate weights per node (\texttt{--numa mirror}). $^\dagger$2 of 5 rounds below 85\,\% of the median.}
\label{tab:runtimes}
\centering
\small
\setlength{\tabcolsep}{3pt}
\resizebox{\columnwidth}{!}{%
\begin{tabular}{@{}lccc@{}}
\toprule
& Llama3-8B & Falcon3-10B & Falcon3-1B \\
\midrule
\ik (tok/s) & 101.5 & 74.2 & 289.8 \\
\ik mirror & 1.029 & 0.902$^\dagger$ & 1.010$^\dagger$ \\
llama.cpp TQ2\_0 & 0.810 & 0.846 & 0.765 \\
llama.cpp TQ2\_0 mirror & 0.739 & 0.801 & 0.735 \\
llama.cpp Q4\_0 & 0.660 & 0.589 & 0.661 \\
llama.cpp Q4\_0 mirror & 0.269 & 0.268 & 0.516 \\
bitnet.cpp I2\_S & 0.776 & -- & 0.814 \\
\bottomrule
\end{tabular}}
\end{table}

\paragraph{ArcLight}
We built ArcLight at commit 558ef06 (May 2026) twice, as shipped for AVX2 and with \texttt{-march=native}, and ran its generator with \texttt{--numa tp} over all four nodes, one thread per physical core (64 or 104), arenas of 12\,GB for weights, 4\,GB for activations, 2\,GB for the KV cache and 2\,GB of work space, a short prompt and 128 generated tokens; we report the decode speed it prints.
Our only patch keeps it from stopping at an end-of-sequence token, so that every run generates 128 tokens.
ArcLight fixes the RoPE base at 10000 instead of reading it from the GGUF file, which changes its output for Llama3 and Falcon3 but not its work, and stores its KV cache as Q4\_0 instead of F16, which is negligible at depth 0.

\section{Barrier isolation on Ice Lake}\label{app:barrier-icl}

\begin{table}[h]
\caption{Barrier isolation on \icl (64 threads, tg128, depth 0, 5 interleaved rounds): per round, \Row's token time and \Col's with 3 promoted or 3 empty global barriers per layer, minus \Col's; median [min, max] over rounds in ms, and the spread of \Col's own token time over the rounds ((max $-$ min) / median).}
\label{tab:barrier-icl}
\centering
\small
\setlength{\tabcolsep}{2pt}
\resizebox{\columnwidth}{!}{%
\begin{tabular}{@{}llcccc@{}}
\toprule
Model & Format & \Row & +3 promoted & +3 empty & \Col spread \\
\midrule
Llama3-8B & IQ2\_BN & 0.52 [0.03, 0.87] & $-$0.44 [$-$0.94, 0.62] & 0.08 [$-$0.59, 0.48] & 11\,\% \\
 & Q4\_0 & 0.88 [$-$0.46, 2.13] & 0.01 [$-$1.19, 1.28] & 1.24 [$-$1.22, 1.48] & 9\,\% \\
Falcon3-10B & IQ2\_BN & 1.07 [0.22, 1.94] & 0.01 [$-$0.69, 0.91] & 0.19 [$-$0.60, 0.29] & 10\,\% \\
 & Q4\_0 & 2.40 [2.18, 2.95] & $-$0.01 [$-$1.01, 1.64] & $-$0.17 [$-$1.04, 1.63] & 7\,\% \\
Falcon3-1B & IQ2\_BN & 0.36 [0.27, 0.43] & $-$0.01 [$-$0.06, 0.05] & 0.06 [0.01, 0.20] & 7\,\% \\
 & Q4\_0 & 0.50 [0.42, 1.16] & 0.11 [$-$0.02, 0.29] & 0.13 [$-$0.04, 0.26] & 4\,\% \\
\bottomrule
\end{tabular}}
\end{table}

We repeated the barrier isolation of the main text on \icl while other users' jobs were running there.
Other processes kept 1.7--3.3 CPUs busy during the runs (we raised the interference limit to 3.5 CPUs for this batch), and since \Col uses every physical core of \icl, they share cores with it: \Col's token time varied by 4--11\,\% between rounds, more than the effects being measured (Table~\ref{tab:barrier-icl}).
\Row is slower than \Col in every configuration, and for Falcon3-1B, where the rounds agree best, three promoted barriers change \Col's token time by $-0.01$\,ms against a gap of 0.36\,ms and three empty ones add 0.06\,ms, so there the barrier count explains a small part of the gap, as on \emr.
For the two larger models the intervals of both variants include zero; these runs do not resolve them.

\section{Accounting of the Row--Col gap}\label{app:gap}

\begin{table}[h]
\caption{\Row minus \Col per-token time (ms) in the batches of the barrier traces (plain runs, medians over 5 rounds), split into the per-stage compute time (barrier waits excluded), the measured time of lm\_head and the leaders' own quantization (LM+Q), and the wait; the wait is divided into waiting for the node leaders, for late threads and the barriers' latency in the proportions of the traced runs. Residual: gap minus all parts.}
\label{tab:gap}
\centering
\small
\setlength{\tabcolsep}{1.6pt}
\resizebox{\columnwidth}{!}{%
\begin{tabular}{@{}llcccccccccc@{}}
\toprule
& & & \multicolumn{5}{c}{compute} & \multicolumn{3}{c}{wait} & \\
\cmidrule(lr){4-8}\cmidrule(lr){9-11}
Model & Format & Gap & QKV & Attn. & O+down & FFN & LM+Q & Leader & Late & Lat. & Resid. \\
\midrule
\multicolumn{12}{@{}l}{\icl, 64 threads} \\
Llama3-8B & IQ2\_BN & 0.66 & 0.41 & 0.03 & $-$0.09 & 0.07 & 0.01 & 0.08 & 0.09 & 0.01 & 0.04 \\
  & Q4\_0 & 0.98 & 0.26 & 0.03 & $-$0.04 & 0.07 & 0.02 & 0.20 & 0.27 & 0.03 & 0.14 \\
Falcon3-10B & IQ2\_BN & 1.00 & 0.62 & 0.05 & $-$0.23 & 0.19 & 0.01 & 0.18 & 0.19 & 0.01 & $-$0.02 \\
  & Q4\_0 & 2.67 & 0.53 & 0.04 & 0.03 & 0.19 & 0.03 & 0.54 & 1.29 & 0.02 & 0.00 \\
Falcon3-1B & IQ2\_BN & 0.43 & 0.23 & 0.01 & $-$0.01 & 0.07 & 0.01 & 0.04 & 0.07 & 0.00 & 0.01 \\
  & Q4\_0 & 0.60 & 0.25 & 0.01 & $-$0.02 & 0.10 & 0.01 & 0.08 & 0.15 & 0.00 & 0.01 \\
\midrule
\multicolumn{12}{@{}l}{\emr, 104 threads} \\
Llama3-8B & IQ2\_BN & 0.83 & 0.15 & 0.15 & $-$0.01 & 0.06 & 0.00 & 0.14 & 0.30 & 0.03 & 0.02 \\
  & Q4\_0 & 1.36 & 0.15 & 0.14 & $-$0.07 & 0.12 & 0.02 & 0.25 & 0.70 & 0.01 & 0.04 \\
Falcon3-10B & IQ2\_BN & 1.25 & 0.26 & 0.08 & $-$0.14 & 0.14 & 0.02 & 0.32 & 0.46 & 0.08 & 0.03 \\
  & Q4\_0 & 1.16 & 0.09 & 0.08 & $-$0.14 & 0.19 & 0.02 & 0.45 & 0.43 & 0.05 & 0.00 \\
Falcon3-1B & IQ2\_BN & 0.50 & 0.13 & 0.01 & 0.00 & 0.01 & 0.00 & 0.05 & 0.25 & 0.03 & 0.01 \\
  & Q4\_0 & 0.61 & 0.13 & 0.02 & $-$0.02 & 0.02 & 0.00 & 0.07 & 0.36 & 0.01 & 0.02 \\
\bottomrule
\end{tabular}}
\end{table}

Within one batch, the measured parts account for the gap to within 0.14\,ms (Table~\ref{tab:gap}).
Two approximations enter: the stage times are means over the whole run, and the wait is divided into its three parts in the proportions measured on the traced runs of the same batch, which cover 32 tokens.
These batches are not the one of the barrier-isolation table in the main text, so their gaps differ from it (e.g., 0.83 instead of 1.01\,ms for Llama3-8B with IQ2\_BN on \emr).
The largest compute difference is the QKV stage, where \Row writes K and V to the node that holds the token's KV block and Q to a shared buffer; the attention stage, which \Row splits by position and \Col by head, differs by up to 0.15\,ms.
We did not isolate these two differences.
\Col's reductions make its O and down stages up to 0.23\,ms longer, and barrier latency contributes at most 0.08\,ms.

\section{KV-cache placement}\label{app:kv}

\begin{table}[h]
\caption{\emr, depth 8192, tg128: share of the KV cache on its most loaded node (0.25 = even over 4 nodes, read from \texttt{numa\_maps}) and decode speed (median, [min, max] of 5 rounds). A memory policy set with \texttt{numactl} takes the process out of automatic NUMA balancing.}
\label{tab:kv}
\centering
\small
\setlength{\tabcolsep}{2.5pt}
\resizebox{\columnwidth}{!}{%
\begin{tabular}{@{}lccc@{}}
\toprule
Variant & KV top share & Llama3-8B & Falcon3-10B \\
\midrule
\ik (default, balancing on) & 0.27--0.54 & 44.2 [42.2, 48.0] & 45.0 [40.2, 48.1] \\
\ik, interleaved KV & 0.25 & 58.1 [57.7, 58.2] & 44.3 [41.1, 44.7] \\
\ik, \texttt{--localalloc} & 1.00 & 32.1 [31.9, 32.2] & 26.1 [26.0, 26.1] \\
\ikp (default, balancing on) & 0.31--0.55 & 45.4 [44.1, 61.9] & 36.3 [28.1, 41.8] \\
\ikp, interleaved KV & 0.25 & 73.8 [73.2, 74.3] & 56.5 [55.8, 57.1] \\
\ikp, \texttt{--localalloc} & 1.00 & 34.2 [34.1, 34.3] & 27.4 [27.3, 27.5] \\
\bottomrule
\end{tabular}}
\end{table}

\ik allocates the KV cache in one buffer on one node (\texttt{--localalloc} keeps it there).
On \emr, automatic NUMA balancing migrates part of it during each run (140--490 thousand pages per run), so the default placement varies from run to run; \icl has balancing off, so there the default is the single-node case.
With the cache interleaved, both \ik and \ikp are stable, and \ikp is 1.27--1.28$\times$ faster than \ik.

\section{Batched decoding}\label{app:batch}

\begin{table}[h]
\caption{Batched decoding (\texttt{llama-batched-bench}, 128-token prompts, 128 tokens per sequence): \ik's total decode throughput (tok/s) and \ikp relative to it. \icl: default KV placement for both. \emr: \ikp with interleaved KV, against the faster of \ik's default and interleaved placement (whose tok/s is shown).}
\label{tab:batch}
\centering
\small
\setlength{\tabcolsep}{2pt}
\resizebox{\columnwidth}{!}{%
\begin{tabular}{@{}r@{\hspace{5pt}}cccccc|cccccc@{}}
\toprule
& \multicolumn{6}{c|}{\icl} & \multicolumn{6}{c}{\emr} \\
& \multicolumn{2}{c}{L-8B} & \multicolumn{2}{c}{F-10B} & \multicolumn{2}{c|}{F-1B} & \multicolumn{2}{c}{L-8B} & \multicolumn{2}{c}{F-10B} & \multicolumn{2}{c}{F-1B} \\
$B$ & \ik & \ikp & \ik & \ikp & \ik & \ikp & \ik & \ikp & \ik & \ikp & \ik & \ikp \\
\midrule
1 & 78.4 & 1.21 & 58.4 & 1.31 & 240.5 & 1.24 & 105.5 & 1.38 & 77.9 & 1.48 & 304.9 & 1.59 \\
2 & 114.2 & 1.05 & 82.3 & 1.09 & 332.9 & 1.19 & 163.5 & 1.17 & 120.8 & 1.21 & 469.8 & 1.33 \\
4 & 195.5 & 1.07 & 99.6 & 1.08 & 434.6 & 1.15 & 237.1 & 1.18 & 170.0 & 1.15 & 719.3 & 1.21 \\
8 & 261.9 & 1.03 & 113.4 & 1.06 & 506.5 & 1.06 & 275.9 & 1.09 & 175.2 & 1.05 & 911.7 & 1.17 \\
16 & 264.6 & 1.06 & 108.0 & 1.02 & 600.5 & 1.02 & 229.1 & 1.04 & 157.6 & 1.04 & 832.0 & 1.05 \\
\bottomrule
\end{tabular}}
\end{table}

On \emr, \ik's default KV placement collapses beyond $B=8$ (Llama3-8B 179.8\,tok/s at $B=8$, Falcon3-10B 102.5 at $B=16$, Falcon3-1B 783.7 at $B=8$); interleaving the cache recovers 18--54\,\%.
Even interleaved, throughput drops from $B=8$ to $B=16$ by 9--17\,\%.

\section{Open observations}\label{app:open}

\begin{itemize}
\item Interleaving the KV cache slows \ik at short context (depth 128) by 15--20\,\% on \icl and 11--13\,\% on \emr.
\item On \icl, Llama3-8B in Q2\_K benefits less from \Col (1.20$\times$) than in Q4\_0 (1.43$\times$), which does not hold on \emr.
\item On \icl, Falcon3-1B has a slightly higher barrier share with \Col (36\,\%) than with \Row (33\,\%), although \Col is 17\,\% faster.
\item \ik's batched throughput on \emr drops from $B=8$ to $B=16$ even with an interleaved KV cache.
\item llama.cpp's weight-replication pull request is 22--59\,\% slower than llama.cpp's own \texttt{distribute} mode with Q4\_0 weights (Table~\ref{tab:runtimes}).
\end{itemize}

\section{AMX prefill and one request}\label{app:amx}

Before the first token, the prompt is processed in one compute-bound matrix--matrix pass.
\emr provides AMX tile instructions with an int8 peak of 552\,Tops, 7.6 times its VNNI peak, which llama.cpp, \ik and bitnet.cpp do not use for ternary weights.
We add an AMX path to \ik's prefill (\amxp).

\paragraph{Ternary GEMM on AMX}
\amxp replaces \ik's IQ2\_BN matrix multiplication whenever at least 32 activation rows are multiplied.
Each thread expands its 2-bit weight rows into the int8 layout of AMX tiles (a 32-bit gather per four input positions of 16 rows, then shift, mask and subtract), unpacks up to 1\,MB of weight panels into L2, and runs all 32-token activation blocks over them; on a socket or more, work is split over both weight rows and tokens.
Activations are quantized exactly as for \ik's VNNI kernel (four scales per row by channel group), with both operands reordered by group so that each 64-wide AMX step stays in one group, so \amxp computes \ik's integer sums.
Its perplexity on Llama3-8B is $0.9994 \pm 0.0014$ of \ik's (KL divergence 0.0022), and its downstream accuracy matches \ik's (Appendix~\ref{app:acc}).
In a microbenchmark with Llama3-8B shapes, the kernel is 4.3--7.2$\times$ faster than the same tiling with VNNI and reaches 50\,Tops on one node and 89 on one socket.

\begin{table}[h]
\caption{Prefill speed (pp512, \emr) of \ik with repacked weights (\texttt{-rtr}, its fastest prefill path) and of \amxp relative to it (median of per-round ratios over 5 interleaved rounds; Falcon3-10B on one socket had 2 low rounds).}
\label{tab:c1}
\centering
\setlength{\tabcolsep}{4pt}
\resizebox{\columnwidth}{!}{%
\begin{tabular}{@{}lcccccc@{}}
\toprule
& \multicolumn{2}{c}{Llama3-8B} & \multicolumn{2}{c}{Falcon3-10B} & \multicolumn{2}{c}{Falcon3-1B} \\
\cmidrule(lr){2-3}\cmidrule(lr){4-5}\cmidrule(l){6-7}
& tok/s & \amxp & tok/s & \amxp & tok/s & \amxp \\
\midrule
1 node (26 thr.) & 524 & 2.07$\times$ & 377 & 1.96$\times$ & 2595 & 1.43$\times$ \\
1 socket (52 thr.) & 965 & 1.79$\times$ & 661 & 1.64$\times$ & 4126 & 1.33$\times$ \\
machine (104 thr.) & 1352 & 1.39$\times$ & 996 & 1.25$\times$ & 3908 & 1.08$\times$ \\
\bottomrule
\end{tabular}}
\end{table}

\paragraph{Prefill throughput}
Against \ik's fastest prefill path, \amxp is 1.43--2.07$\times$ faster on one node, 1.33--1.79$\times$ on one socket and 1.08--1.39$\times$ on the whole machine (Table~\ref{tab:c1}).
The gain shrinks with scale (from one node to the machine, \ik's prefill grows 2.6$\times$ on Llama3-8B and \amxp's 1.7$\times$): once the matrix products run on AMX, the remaining operators and ggml's per-operator synchronization take a larger share; we did not apply the engine's dataflow to prefill.
llama.cpp's Q4\_0 path, which uses AMX after repacking, reaches 0.69--0.78$\times$ of \ik's fastest prefill on one node or socket (2.4--3.0$\times$ its own speed without AMX); llama.cpp TQ2\_0 and bitnet.cpp I2\_S reach 0.36--0.56$\times$.
Intel's TPP 2-bit GEMM, which also uses AMX, ran the prefill of Ternary-Bonsai-8B about 1.8$\times$ (one node) and 1.1$\times$ (one socket) faster than \ik on the same model but decoded at 0.31--0.37$\times$ of \ik's speed; this comparison uses a different model and framework and was not interleaved.
In batched decoding with 32 or 64 sequences, where decoding becomes a matrix product, \amxp is 5--21\,\% faster than \ik on one node and within noise on the whole machine.

\begin{figure*}[t]
\centering
\includegraphics[width=\textwidth]{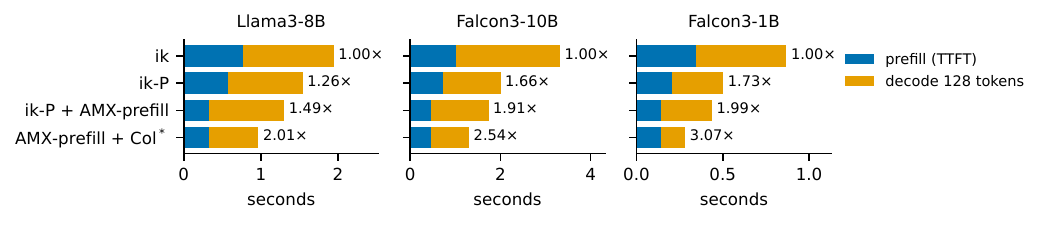}
\caption{One request on \emr (104 threads): 512-token prompt, 128 generated tokens; prefill and decode time (medians of 5 rounds) and speedup over \ik. $^*$\amxp prefill plus \Col decode at depth 512, measured separately with different timers.}
\label{fig:e2e}
\end{figure*}

\paragraph{One request}
For a request with a 512-token prompt and 128 generated tokens on the whole \emr machine (Fig.~\ref{fig:e2e}), \amxp shortens the time to first token to 1/2.2--1/2.4 of \ik's.
In one end-to-end run, \ikp with \amxp completes the request 1.49--1.99$\times$ faster than \ik.
Adding \Col's separately measured decode time to the prefill time of \amxp gives 2.01--3.07$\times$; this is an estimate from two phases timed by different tools, as the engine has no prefill of its own.
Decoding still takes 50--66\,\% of the request with \Col.

\end{document}